\documentclass[aps,prc,twocolumn,superscriptaddress,10pt,english,preprintnumbers]{revtex4-1}
\pdfoutput=1
\usepackage[T1]{fontenc}
\usepackage[utf8]{inputenc}
\usepackage{epsfig}
\usepackage{graphicx}
\usepackage{xcolor}
\usepackage{latexsym}
\usepackage{textcomp}
\usepackage{amssymb}
\usepackage[colorlinks=true,linkcolor=blue,citecolor=blue, urlcolor=blue]{hyperref} 
\usepackage{amsfonts,amsthm,amstext,amscd}
\usepackage{amsmath}
\usepackage{mathtools}
\usepackage{comment}

\newcommand{\tr}{\emph{Trajectum}}
\DeclareMathOperator{\sgn}{sgn}

\begin{document}
\title{Hard probe path lengths and event-shape engineering of the quark-gluon plasma}
\author{Caitlin Beattie}
\affiliation{Physics Department, Yale University, New Haven, CT 06511, USA}
\author{Govert Nijs}
\affiliation{Center for Theoretical Physics, Massachusetts Institute of Technology, Cambridge, MA 02139, USA}
\author{Mike Sas}
\affiliation{Physics Department, Yale University, New Haven, CT 06511, USA}
\affiliation{European Organization for Nuclear Research (CERN), Geneva, Switzerland}
\author{Wilke van der Schee}
\affiliation{Theoretical Physics Department, CERN, CH-1211 Gen\`eve 23, Switzerland}
\begin{abstract}
As particles traverse the quark-gluon plasma (QGP) formed during a heavy ion collision they undergo energy loss depending on the distance travelled. We study several temperature- and velocity-weighted path length distributions of non-interacting particles as they traverse the plasma using the \emph{Trajectum} heavy ion code, including those of back-to-back path lengths. We use event-shape engineering (ESE) in combination with in-plane versus out-of-plane selection to accurately control these path lengths. Lastly, we show how soft observables depend on the different ESE classes.
\end{abstract}

\preprint{CERN-TH-2022-051//MIT-CTP/5397}

\maketitle

\section{Introduction}

Ultrarelativistic heavy ion collisions can create a small droplet of deconfined quark matter in the form of strongly coupled quark-gluon plasma (QGP) that allows for the study of quantum chromodynamics (QCD) in its non-perturbative regime. At the same time, energetic quarks and gluons (hard partons) are created by perturbative processes. The study of the QGP typically involves analyses of soft (low $p_{T}$) particles whose behavior can be described by hydrodynamical models, as well as those hard partons that can interact with the QGP\@. The ability to bridge these two regimes offers promise for understanding the interactions between the hard and soft scales. Typical observables for such interactions are the reduction in yields of hard particles or sprays of particles (jets, \cite{1802.04801,1805.05635,1909.09718}), or the anisotropy of these particles or jets with respect to the soft particles \cite{1802.04801,1702.00630}, both of which contain valuable information on, for instance, the in-medium parton energy loss \cite{1802.04801,1202.5022,1805.05635}\@.

\begin{figure}[t]
\includegraphics[width=0.45\textwidth]{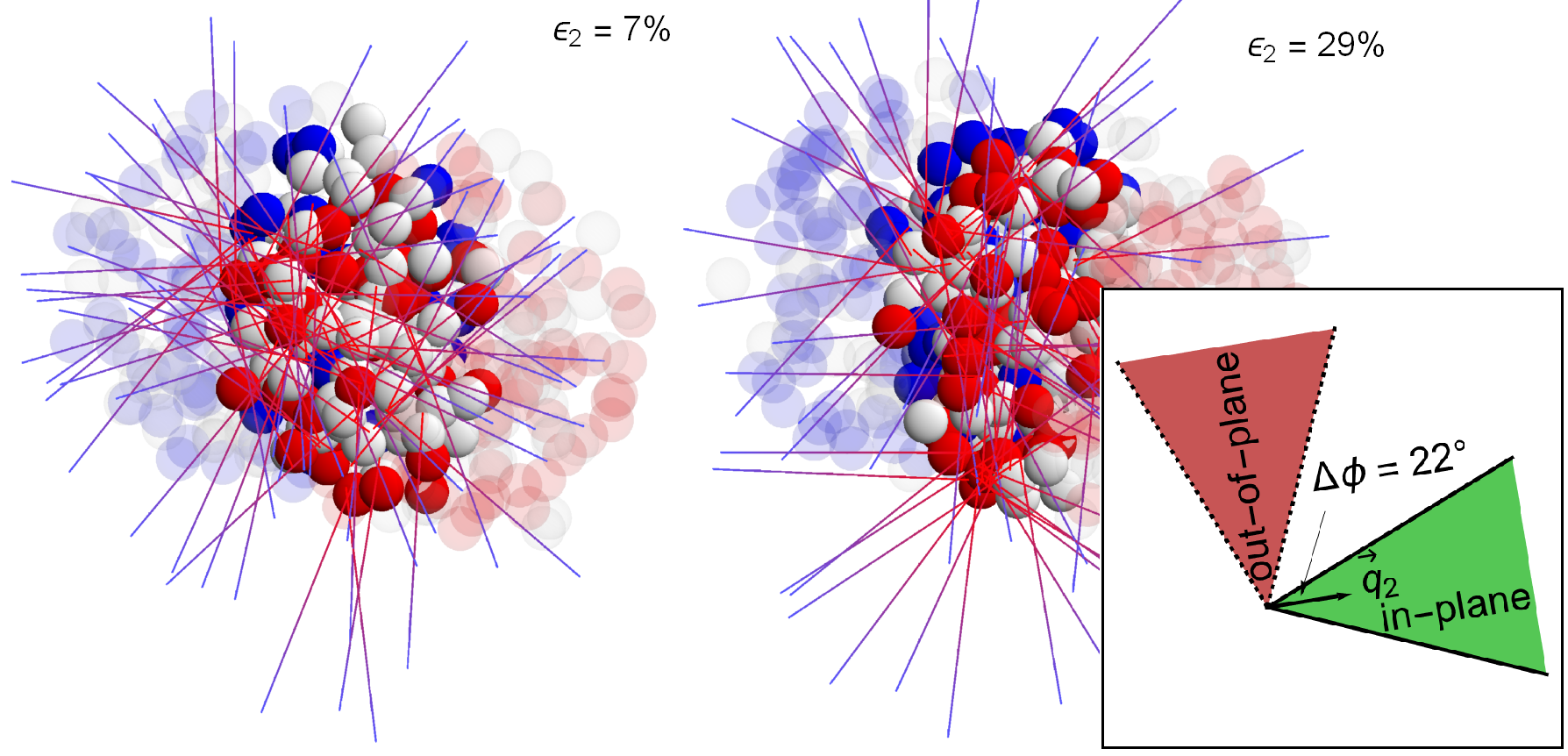}
\caption{\label{fig:illustration}An illustration of two collisions of two Pb nuclei with impact parameter $b=7\,$fm (approximately 25\% central)\@. Paths of hard partons are distributed in a random direction at the locations of each binary nucleon-nucleon collision (only 10\% shown as red to blue rays of $8\,$fm)\@. The initial eccentricities $\epsilon_2$ are different in both cases, showing the different geometries that are possible at the same centrality. Paths are considered to be in-plane by having a difference in azimuthal angle with respect to the $Q_2$ vector \eqref{eq:Qn} of less than $22^\circ$ as illustrated, with an analogous definition for out-of-plane paths.}
\end{figure}

In this Letter we explore the bridge between hard and soft observables by studying paths traversed by hard partons in connection with the shape of the background QGP\@. A more elliptically shaped initial QGP will give rise to a more anisotropic distribution of soft particles \cite{1802.04801,1105.3865,1212.1008,1505.02677,2010.15134} and at the same time will feature characteristically different path lengths in the short versus long direction (see Fig.~\ref{fig:illustration})\@. This can be quantified by the technique of event-shape engineering (ESE) \cite{1208.4563}, which classifies events within a centrality class by their anisotropies and can hence select on the shape of the initial QGP while keeping other properties similar. This is also illustrated in Fig.~\ref{fig:illustration}, which shows two events of roughly the same centrality, while the rightmost event has a much larger eccentricity $\epsilon_2$\@. In this Letter we will apply ESE within the \tr{} heavy ion code \cite{2010.15134, 2010.15130, 2206.13522} (code available at \cite{trajectumcode}), giving particular attention to connect processes at hard and soft scales. We calculate the path length distributions that a hard parton would traverse as a function of centrality and ESE selection and also consider how this path length varies with respect to the event-plane angle of the collision. For a realistic modelling of parton energy loss, we further consider entropy- and fluid velocity-weighted paths, both for the event-plane angle selected paths as well as for back-to-back paths. Finally, we show soft observables as computed by \tr{} for several ESE classes.

\begin{figure*}[ht]
\includegraphics[width=0.4\textwidth]{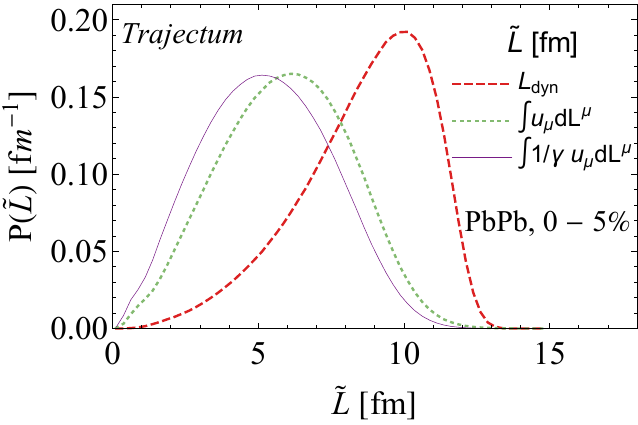}
\includegraphics[width=0.38\textwidth]{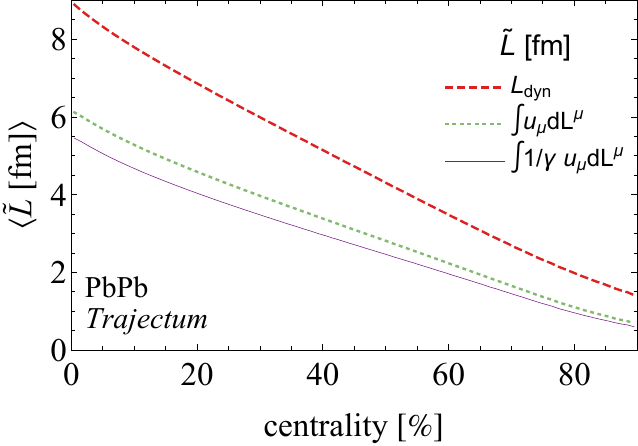}
\caption{\label{fig:differentpathlengths}Path length distributions of non-interacting probes traversing the plasma for 0--5\% collision centrality (left) and the mean path length as a function of collision centrality (right)\@. The four different approaches to calculate $\tilde L$ are described in the method section.}
\end{figure*}

Experimentally there have been several studies of parton energy loss for both in-plane and out-of-plane directions as a function of centrality both at RHIC \cite{STAR:2004amg,PHENIX:2006wwy,1010.0690} and the LHC \cite{1509.07334,1910.14398,2111.06606}\@. In general energy loss decreases towards higher centralities as the QGP becomes smaller. The difference between in- and out-of-plane probes increases, since towards higher centralities the anisotropy increases. The main strength of ESE done here becomes clear: it is possible to vary the anisotropy while keeping the QGP size approximately constant (see also \cite{1606.07963})\@.

We should note that the present study is quite limited if one would like to compare with experiments. An important contribution here is that we only compute (effective) path lengths, which are not directly related to parton energy loss. In particular, for energy loss of jets it is often crucial that even for fixed paths there is a wide energy loss distribution. Together with the steeply falling spectrum of jets this leads to a measured sample that contains mostly jets that lost relatively little energy \cite{1512.08107, 1602.04187,1710.03237}\@. It would hence be interesting to see how such fluctuations would influence our results. This is also particularly relevant for dijets, where it has been shown that energy loss distributions due to fluctuations are as important as path length distributions \cite{1512.08107,1601.03629, 1809.10695}\@. The latter conclusion is supported by either turning off path length variations entirely (by starting dijets in the center of a spherical plasma \cite{1512.08107,1809.10695}) or by turning of fluctuations while varying the dijet path lengths \cite{1809.10695}. Surprisingly, both counter-factual set-ups result in the same dijet asymmetry distribution, confirming that both fluctuations and path length asymmetry are as important to understand the dijet assymetry distribution. We note, however, that the set-up with all dijets starting in the center overestimates the energy loss and hence does not describe the nuclear modification factor for inclusive jets \cite{1809.10695}.

It is important that for this study the model for the soft sector is fitted to experimental data, meaning that the free parameters in \emph{Trajectum} reproduce state-of-the-art geometries for the QGP resulting from a collision \cite{2010.15130,2010.15134,2110.13153,2206.13522}\@.
In particular, event-by-event fluctuations are incorporated into the initial state which goes into the hydrodynamics model.
Fluctuations of the initial state are important for various observables.
For example, triangular flow $v_3$ is entirely generated through fluctuations, since by symmetry it should otherwise be equal to zero.
For our study of path length distributions of hard probes, we similarly expect that fluctuations will play a large role, as they have a large effect on the shape of the plasma which is being traversed by the hard probe.
The presence of fluctuations also gives experiments a wider range of observables, as ESE and centrality selections allow for the shape and size of the QGP to be independently varied.

\section{Model and path length measures}
In this work, we use \emph{Trajectum} version 1.2 to generate 2.2M hydrodynamic events using the maximum likelihood (MAP) settings from \cite{2206.13522} at a center-of-mass energy of $\sqrt{s_{\mathrm{NN}}}=5.02\,$TeV\@. In particular, this means that we generate boost invariant initial conditions using a modified version of the T\raisebox{-0.5ex}{R}ENTo model, and then use second order hydrodynamics to evolve the produced QGP\@.
We then find the freeze-out hypersurface defined as the isotemperature surface where $T = 150.1\,\text{MeV}$\@.
At the freeze-out surface we apply the Cooper-Frye formula \cite{Cooper:1974mv} to go from a fluid description to a particle prescription, with viscous corrections given by the Pratt-Torrieri-Bernhard (PTB) prescription \cite{1003.0413, 1804.06469}\@.
Lastly, we feed the resulting particles into SMASH \cite{1606.06642} to include final state interactions.

The path length that is traversed in the quark-gluon plasma by a non-interacting probe depends on its origin and direction, as well as the size and expansion of the plasma. A probe created near the surface of the plasma that is directed outward will have a shorter path length compared to one that is created around the center. In the most extreme example, a probe can be created near the surface and directed towards the center, essentially traversing the entire size of the medium along one axis.

We model the distance a hard parton would traverse by generating lines through the plasmas produced in the events. These lines are produced in pairs at the location of each nucleon-nucleon collision (see Fig.~\ref{fig:illustration}, only one of the back-to-back probes shown to avoid cluttering), and propagate outwards in opposite directions at midrapidity until they encounter a temperature smaller than the freeze-out temperature. Given that the plasma is boost-invariant, we can choose the lines to propagate at rapidity $y=0$ without loss of generality.

In the simplest setting we compute the path length from the starting point of each line to the freeze-out surface. Next, we also compute the integrals of several quantities over these paths, which can potentially estimate the energy loss of a hard probe (see also \cite{Dainese:2004te})\@. Note here that, since fluid properties such as temperature are not well-defined before $\tau_\text{hyd} = 0.370\,\text{fm}/c$, the integrand is taken to be zero before $\tau_\text{hyd}$\@. 

\begin{figure*}[t]
\includegraphics[width=0.65\columnwidth]{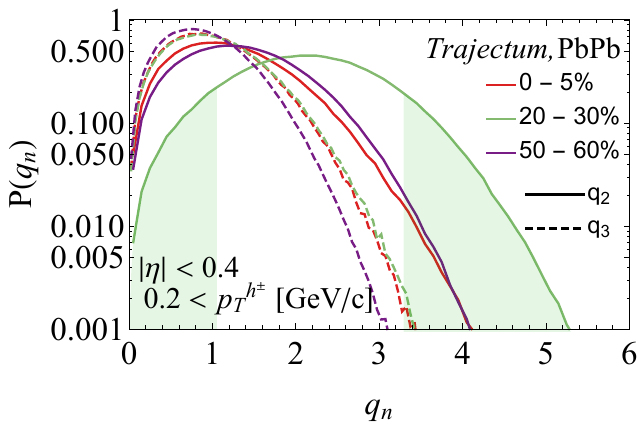}
\includegraphics[width=0.65\columnwidth]{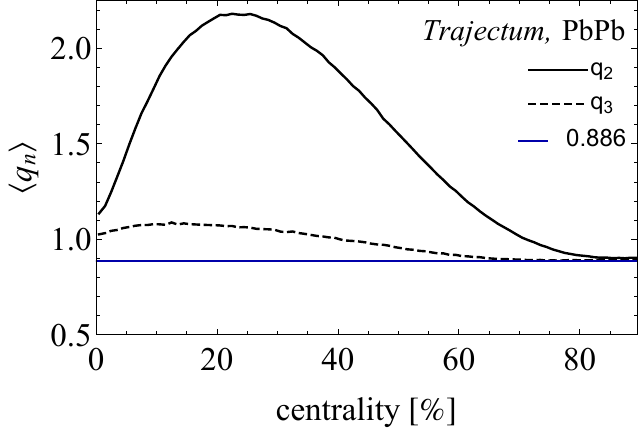}
\includegraphics[width=0.65\columnwidth]{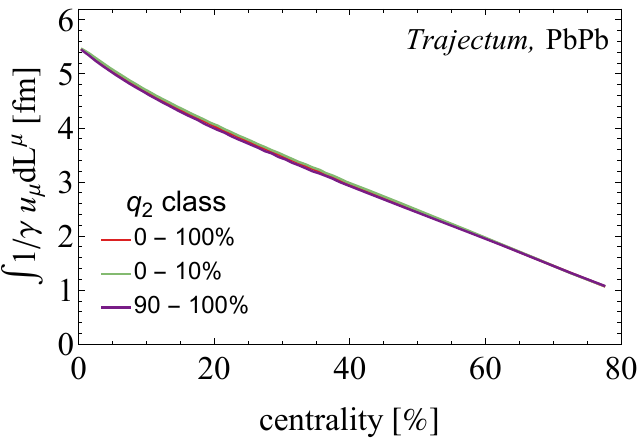}
\caption{\label{fig:LwithESE}Left: Distribution of $q_n$ values for different centralities, using the same track selections as in \cite{1507.06194} ($p_T > 0.2\,$GeV and $|\eta|<0.4$)\@. Also shown for 20--30\% centrality are the 10\% lowest and the 10\% highest $q_2$ values (shaded)\@. Middle: average values of the $q_{n}$ distributions. Right: Average path length $\langle\int 1/\gamma\ u_\mu dL^\mu\rangle$ of non-interacting probes traversing the plasma as a function of collision centrality for different selections on the ESE observable $q_{2}$\@.}
\end{figure*}

A naive estimate of the energy loss of a parton would integrate a power $\alpha$ of the local temperature $T^\alpha$, where in this work we use $\alpha=3$ under the assumption that energy loss scales with the entropy density \cite{0801.2173,1312.5003,1507.06556,2007.13758,2007.10078}\@. Realistically, however, both the parton and the fluid move relativistically and $T^\alpha$ should only be used in the fluid rest frame. To capture this effect we integrate
\begin{equation}
dE = \frac{T^\alpha}{\gamma}u_\mu dL^\mu, \label{eq:energyloss}
\end{equation}
with $\gamma$ the Lorentz factor, $u^\mu$ the local fluid velocity and $dL^\mu$ a length element \cite{Baier:2006pt}. Such a description is correct for probes propagating parallel to the fluid. In this case $u_\mu dL^\mu/\gamma = (\gamma-v\gamma)/\gamma = 1 - v$, with $v$ the fluid velocity. This factor reduces energy loss when travelling with the flow and enhances it for countercurrent propagation. The factor $1-v$ agrees with classical intuition that it is not necessarily the parton velocities themselves that matter (after all these energies are small compared to the probe energy), but it is the geometric effect that per unit time the number of partons encountered increases or decreases by a factor $1-v$\@.
For probes propagating perpendicular to the fluid, \eqref{eq:energyloss} is correct up to $\mathcal{O}(|\vec{v}|^2)$\@. We note that such perpendicular flows induce momentum in the fluid flow direction $\propto T^\alpha |\vec{v}|^2$ and can hence change the direction of the probe, but in this work we neglect this effect.

In the following we show distributions obtained from integrating \eqref{eq:energyloss}, but we stress that in this work we do not actually compute energy loss as \eqref{eq:energyloss} is only proportional to our ansatz for energy loss. Nevertheless, as a proxy for energy loss (up to a constant factor) the results can be used to identify interesting regimes for ESE\@.

It is important that path lengths are evaluated from the origin of the line to the point where it leaves the plasma, but that the integrals are evaluated only from the time $\tau_\text{hyd}$ onward, since the required quantities of $T$ and $u^\mu$ are unknown before this time. Each of the elements in \eqref{eq:energyloss} can be separately included in the computation, i.e.~we can choose the power $\alpha$ while independently omitting or including the factor $\gamma$ and the inner product with the velocity for the purpose of evaluating each element's contribution to the energy loss.

Fig.~\ref{fig:differentpathlengths} shows the distributions of path length $\tilde L$ for 0--5\% collision centrality (left) and the average $\langle\tilde L\rangle$ for all collision centralities (right) for $\tilde L$ equal to the path lengths as well as two integrals taking into account fluid flow (see legend)\@. 
The path lengths have a sharp cut-off around $\tilde L=12\,\text{fm}$ since after this time there is no QGP left. 
For the more realistic cases with fluid velocities we see the path lengths are strongly reduced, which implies that most hard probes propagate in the direction of the fluid. An extra factor of $1/\gamma$ reduces this further by about an extra 10\%\@.

\begin{figure*}[ht]
\includegraphics[width=0.99\textwidth]{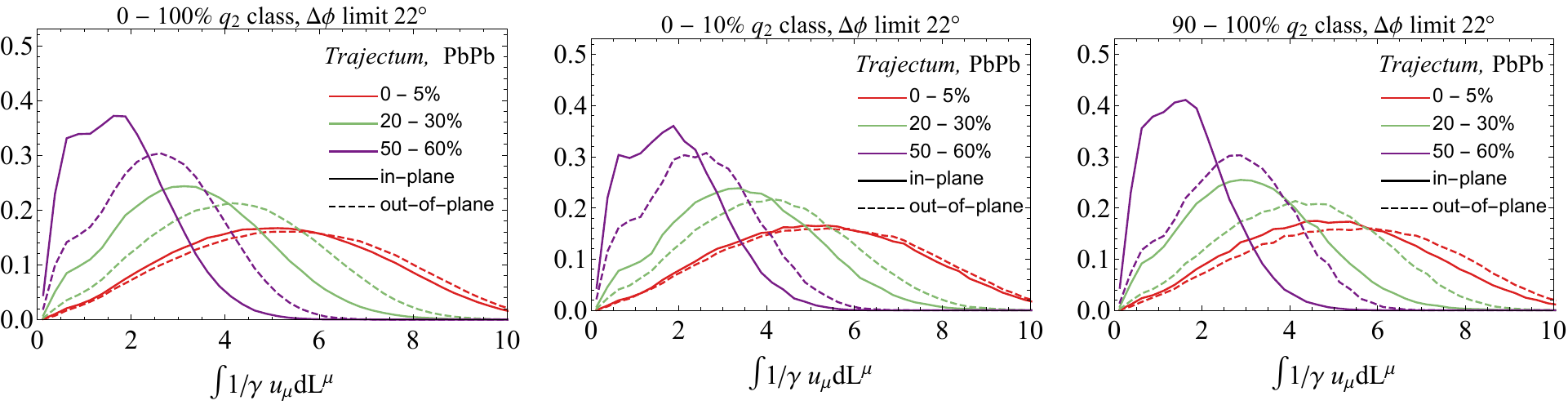}
\caption{\label{fig:pathlengthsinout}In- and out-of-plane path length distributions of non-interacting probes traversing the plasma, for three different collision centralities. The three panels differ in their selection on the ESE-observable $q_{2}$ with 0--100\% (left), 0--10\% (middle), and 90--100\% (right)\@.}
\end{figure*}

\begin{figure*}[ht]
\includegraphics[width=0.99\textwidth]{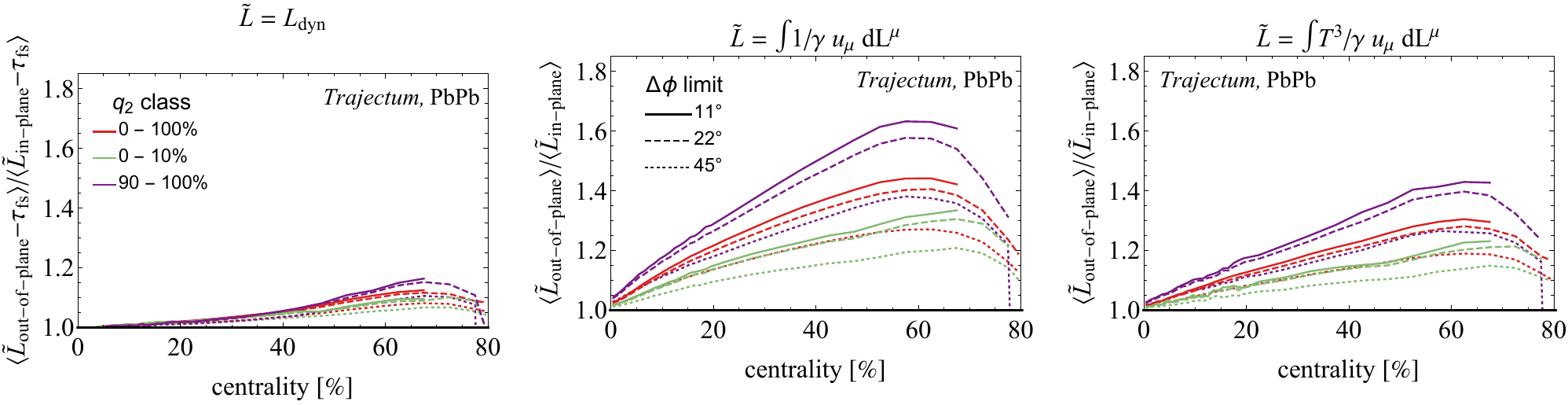}
\caption{\label{fig:inoutratio}Ratio of the mean in- and out-of-plane path lengths as a function of collision centrality, for three different $q_{2}$ classes as well as for $\Delta \phi$ emission angle limits of $11^\circ$ (solid), $22^\circ$ (dashed), and $45^\circ$ (dotted)\@. The three panels correspond to path lengths, fluid-flow weighted path lengths and temperature plus fluid flow-weighted path lengths, respectively.}
\end{figure*}

\begin{figure*}[ht]
\includegraphics[width=0.99\textwidth]{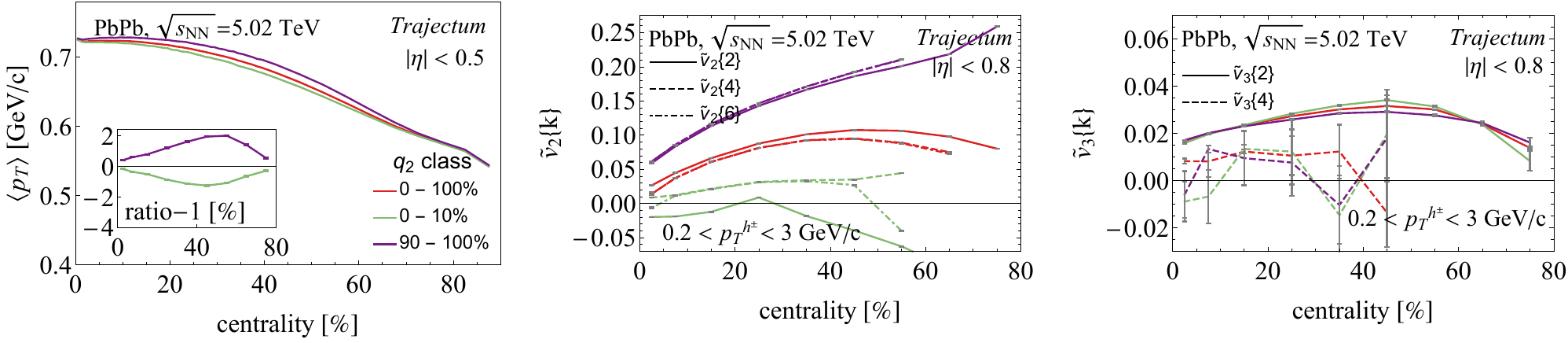}
\includegraphics[width=0.99\textwidth]{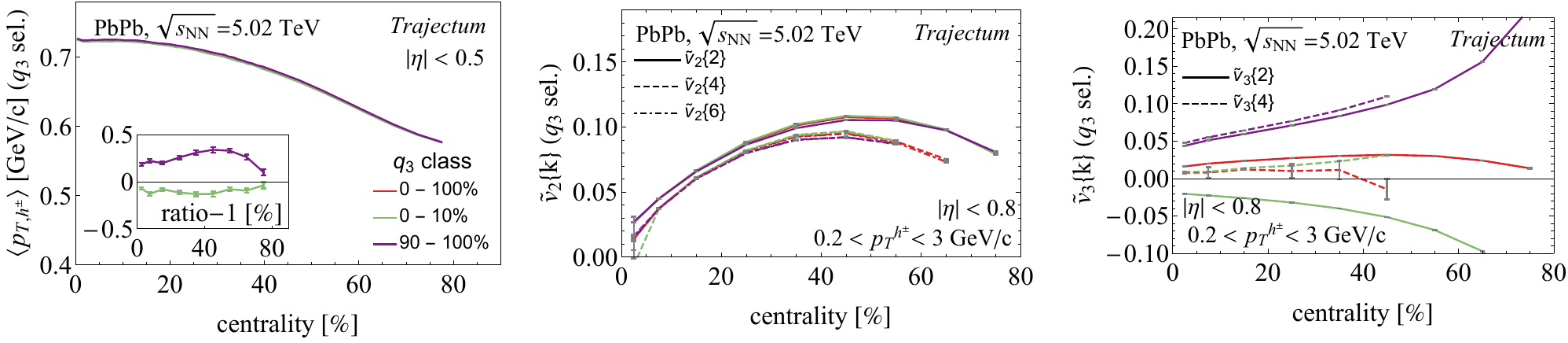}
\caption{\label{fig:observables}We show mean transverse momentum $\langle p_T\rangle$ (left), elliptic flow $\tilde v_2\{k\}$ (middle) and triangular flow $\tilde v_3\{k\}$ (right), as a function of centrality and using ESE selection on $q_2$ (top) and $q_3$ (bottom)\@. Shown as insets in the left panels are the ratios of the 0--10\% and 90--100\% $q_n$ classes with the 0--100\% $q_n$ class.}
\end{figure*}

Many of the results shown in the rest of this work use event shape engineering (ESE) to select particularly elliptical or particularly spherical events.
This is achieved through the flow vector defined by
\begin{equation}\label{eq:Qn}
Q_n=\sum_{i=1}^{M} e^{in\varphi_i},
\end{equation}
where $M$ is the particle multiplicity and $\varphi_{i}$ the azimuthal angle of particle $i$\@. The sum is performed over charged particles with transverse momentum $p_T > 0.4\,\text{GeV}/c$ and pseudorapidity $|\eta| < 2.4$\@. 
The flow vector is related to the reduced flow vector $q_n$ and the two-particle anisotropic flow observable $v_n\{2\}$ through
\begin{equation}\label{eq:Qnv2}
q_{n} = \frac{|Q_{n}|}{\sqrt{M}}, \quad v_n\{2\}^2 = \left\langle \frac{q_n^2 - 1}{M-1}\right\rangle.
\end{equation}
Here the averages are performed over events within the class of events considered. The reduced flow vector is defined in such a way that if the $\varphi_i$ are completely uncorrelated it would equal a constant value of $\langle q_n \rangle = 0.886$ for large multiplicity (note that in such a case $\langle q_n^2 \rangle=1$ as in the 2D random walk problem)\@. For larger values there is a non-trivial two-particle correlation, as expected from hydrodynamic evolution. Fig.~\ref{fig:LwithESE} (left) shows the distribution of $q_2$ and $q_3$ values for different centralities, using the same track selections as in \cite{1507.06194}\@. For the $q_2$ distribution in the 20--30\% centrality class, we also show as shaded regions the lowest (0--10\%) and highest (90--100\%) $q_2$ values\@. These shaded regions correspond to the low and high ellipticity selections that we use in figures where we use ESE\@. Also shown in Fig.~\ref{fig:LwithESE} (middle) is the average $q_n$ as a function of centrality, which indeed asymptotes to 0.886 for very peripheral events. Quite remarkably $q_2$ does not depend as strongly on centrality as $v_2\{2\}$ \cite{1507.06194}, since both $v_2\{2\}$ (see also Fig.~\ref{fig:observables} later) and $1/\sqrt{M}$ increase as a function of centrality up to centralities of around 50\%\@. It is for this reason that $q_2$ is a good quantity for ESE (as opposed to for instance $v_2\{2\}$), which groups collision events according to centrality (multiplicity) as well as $q_n$ within such a class.

Even with the moderate centrality dependence it means that selecting, for example, the 10\% most elliptical events in a large centrality bin can unintentionally bias the sample towards a large (or small) centrality within that bin. When this occurs, the ESE selection on $q_2$ becomes an effective selection on centrality, thus prohibiting any additional gain in information from the ESE selection. To avoid this problem, we always compute observables in 1\% wide centrality bins (using ESE within these small bins when relevant), and then average the results over wider centrality bins. The use of 1\% wide centrality bins proves sufficiently narrow to avoid the problem described above, and also coincides with the prescription that experiments follow \cite{1507.06194,1809.09371}\@.

\begin{figure}[h]
\includegraphics[width=0.79\columnwidth]{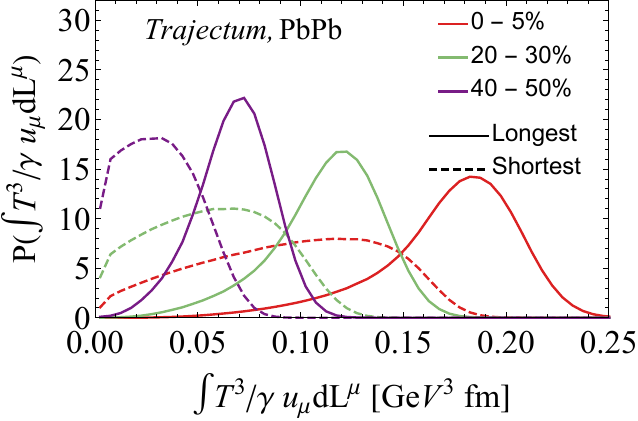}
\includegraphics[width=0.79\columnwidth]{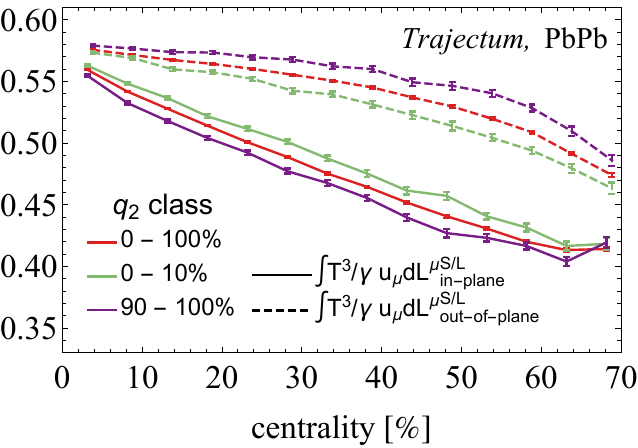}
\caption{\label{fig:dijets}Top: path length distributions of non-interacting back-to-back probes, separating the longest and shortest paths for the 0--5\%, 20--30\%, and 40--50\% collision centralities. Bottom: ratio of the shortest over the longest path for each dijet, for in- (solid) and out-of-plane (dashed, both using a $\Delta\phi$ emission angle limit of  $22^\circ$), as a function of collision centrality for the different $q_{2}$ classes. Note the short-hand $S/L$ notation which really divides over the entire integrals of the short and long paths.}
\end{figure}

Fig.~\ref{fig:LwithESE} (right) shows the average path length $\langle\int1/\gamma\ u_\mu dL^\mu\rangle$ of non-interacting probes traversing the plasma as a function of collision centrality for an inclusive sample, denoted by 0--100\%, as well as for the percentiles of lowest (0--10\%) and highest (90--100\%) values of $q_{2}$\@. It shows that within the model there is no meaningful difference in traversed path lengths when selecting events based on $q_{2}$ for any of the collision centralities. This is attributed to the fact that for events with higher eccentricity some paths will shorten (in-plane) and some will lengthen (out-of-plane), mostly averaging out when looking at all possible emission angles.

Fig.~\ref{fig:pathlengthsinout} shows the path length distributions separately for in- and out-of-plane emission angles for three different collision centralities as well as three selections on the ESE observable $q_{2}$\@. Here, in-plane (out-of-plane) is defined by including paths that are within an azimuthal angle difference of $\Delta \phi = 22^\circ$ with respect to the short (long) axis, as extracted from the event plane angle defined by the argument of $Q_2$ (see also Fig.~\ref{fig:illustration})\@. Note here, that this argument of $Q_2$ differs from the participant plane due to the finite number of particles in the final state \cite{nucl-th/9711003,nucl-ex/9805001,nucl-ex/0311007}\@. In principle one should expect to see the strongest effects when selecting with respect to the participant plane. However, given that the experiments do not have access to the participant plane, we make the choice to use the quantity that can be measured experimentally. The three panels show the results of all $q_{2}$ values (left), the 0--10\% percentile of $q_{2}$ (middle), and the 90--100\% percentile (right) of $q_{2}$\@. 
As expected, path lengths are typically larger for central collisions or for paths in the out-of-plane direction, since in those cases there is more plasma to encounter. When selecting on $q_2$ (middle and right) the difference between in- and out-of-plane becomes significantly smaller (middle, small $q_{2}$) or larger (right, large $q_2$), whereby this effect is much larger for semi-central or peripheral collisions. Indeed, central collisions produce a system with small eccentricities which then reduces the dependence on emission angles. There is some structure in Fig.~\ref{fig:pathlengthsinout} around lengths from $0.5$ to $2\,\text{fm}$, which is not due to statistical uncertainty, as this uncertainty is negligibly small. This structure could possibly be due to the lumpiness in the initial state caused by the finite size of individual nucleons.

The ratio of the mean out-of-plane to in-plane path lengths, $\langle L_{\text{out-of-plane}} \rangle /  \langle L_{\text{in-plane}} \rangle$, is calculated as a function of collision centrality for three different $q_{2}$ classes, with the results shown in Fig.~\ref{fig:inoutratio}\@. Here, we include the length measure $\tilde L = \int T^{3}/\gamma\ u_\mu dL^\mu$, in addition to choices shown previously, as energy loss is expected to go with $T^3$ \cite{0801.2173,2007.10078}\@. The calculations are performed for the $\Delta \phi$ emission angle limits of $11^\circ$, $22^\circ$, and $45^\circ$, where $\Delta \phi$ is the azimuthal angle difference between the probe and the in- and out-of-plane axes. Without a selection on $q_{2}$ it is possible to achieve an unweighted in- and out-of-plane path length ratio of only 1--1.2 for the $\Delta \phi=45^\circ$ limit, increasing from central to semi-central collision centrality. This effect can be enhanced by decreasing the $\Delta \phi$ limit to $22^\circ$ or $11^\circ$\@. Here it is interesting to note that the impact of going from $\Delta \phi = 45^\circ$ to $\Delta \phi = 22^\circ$ is much larger than further restricting it to $\Delta \phi = 11^\circ$\@. As such, it might be best for experimental measurements to use $\Delta \phi = 22^\circ$, which nearly doubles the differences in path lengths, while $\Delta \phi = 11^\circ$ would mainly reduce the available statistics. Furthermore, the largest modification of path lengths can be accessed by making an additional selection on the ESE observable $q_{2}$\@. Compared to the 0--100\% $q_{2}$ class, the collisions with low $q_{2}$ show nearly no change for in- and out-of-plane path lengths, while the collisions with high $q_{2}$ are further enhanced by almost a factor two (purple versus red)\@. It is interesting that all these effects are much larger for flow-weighted paths (middle and right) as opposed to just integrating the length of the paths.
Since this difference is possibly due to the fact that the flow-weighted paths are integrated from $\tau = \tau_\text{hyd}$ onward instead of from $\tau = 0$ as is done for $L_\text{dyn}$, for $L_\text{dyn}$ we subtract $\tau_\text{hyd}$ for both the in-plane and out-of-plane averages. This does not have an effect on our conclusions.
We therefore conclude that when determining energy loss, consideration of the flow of the fluid is as important as consideration of path lengths.

Another way to study energy loss is via back-to-back dijets, where typically one of the jets will traverse more medium than the other. In the model we generate back-to-back probes and keep track of the traversed length as the usual $\tilde L = \int T^{3}/ \gamma\ u_\mu dL^\mu$, and separate the longest from the shortest trajectories through the plasma. The resulting distributions of these effective path lengths are, for the 0--5\%, 20--30\%, and 40--50\% collision centralities, shown in Fig.~\ref{fig:dijets} (top)\@. It shows that the longest path length is roughly twice as large as the shortest for central collisions, with shorter path lengths for more peripheral collisions. Furthermore, we calculate the ratio of the shortest over the longest path for each dijet that is created in- and out-of-plane, as a function of collision centrality for the different $q_{2}$ classes. This ratio is shown in Fig.~\ref{fig:dijets} (bottom)\@. For the $q_{2}$ integrated case (0--100\%), the $S/L$ ratio for in-plane back-to-back path lengths is about 15\% below the out-of-plane back-to-back path lengths for semi-central collision centralities. This difference gets enhanced again when only selecting events with low or high values of $q_{2}$ as we also saw in previous results. This implies that in-plane back-to-back path lengths traverse significantly more unbalanced effective path lengths, with an increasing effect going towards more peripheral collisions and selecting on high $q_2$ events.

\section{Bulk observables}
In addition to the path length observables, one can also apply ESE to some of the bulk observables. Fig.~\ref{fig:observables} shows the mean transverse momentum $\langle p_T\rangle$, as well as $\tilde v_2\{k\}$ and $\tilde v_3\{k\}$, using ESE selection on both $q_2$ and $q_3$\@. Note here that by definition $\tilde v_n\{k\} = \sgn(v_n\{k\}^k)|v_n\{k\}|$ so that $\tilde v_n\{k\}$ is never complex, but instead takes on negative values for complex $v_n\{k\}$ \cite{2010.15134}\@. The $v_n\{k\}$ themselves are defined through cumulants as follows. For $v_n\{2\}$ we have
\[
v_n\{2\} = \sqrt{\langle\langle 2_n\rangle\rangle}, \qquad \langle 2_n\rangle = \langle e^{in(\varphi_i-\varphi_j)}\rangle_{i\neq j},
\]
where the inner $\langle\cdot\rangle$ is an average over particle pairs within each event, and the outer $\langle\cdot\rangle$ averages over all events in a centrality and ESE class, where each event is weighted by the number of terms in $\langle 2_n\rangle$\@.
For $v_n\{4\}$ and $v_n\{6\}$, we have
\begin{align*}
v_n\{4\} & = \left(-\langle\langle 4_n\rangle\rangle + 2\langle\langle 2_n\rangle\rangle^2\right)^{1/4}, \\
v_n\{6\} & = \left(\frac{1}{4}(\langle\langle 6_n\rangle\rangle - 9\langle\langle 2_n\rangle\rangle\langle\langle 4_n\rangle\rangle + 12\langle\langle 2_n\rangle\rangle^3)\right)^{1/6},
\end{align*}
where $\langle 4_n\rangle$ and $\langle 6_n\rangle$ are defined in analogy to $\langle 2_n\rangle$ by
\begin{align*}
\langle 4_n\rangle & = \langle e^{in(\varphi_i+\varphi_j-\varphi_k-\varphi_l)}\rangle_{i\neq j\neq k\neq l}, \\
\langle 6_n\rangle & = \langle e^{in(\varphi_i+\varphi_j+\varphi_k-\varphi_l-\varphi_m-\varphi_n)}\rangle_{i\neq j\neq k\neq l\neq m\neq n},
\end{align*}
where all the indices are unequal.
All the $\langle k_n\rangle$ can be evaluated in terms of $Q_n$\@.
The expressions for $\langle 4_n\rangle$ and $\langle 6_n\rangle$ are fairly lengthy \cite{1010.0233,Bilandzic:2012wva}, but the expression for $\langle 2_n\rangle$ is shown in \eqref{eq:Qnv2}\@.
The first basic thing one can check is what happens to $\tilde v_2\{2\}$ and $\tilde v_3\{2\}$ when selecting for $q_2$ and $q_3$, respectively.
As expected, selecting the events with the largest (smallest) $q_2$ increases (decreases) $\tilde v_2\{2\}$, indicating that indeed ESE is giving us a handle to select events with large or small ellipticity.
Similarly, demanding large (small) $q_3$ increases (decreases) $\tilde v_3\{2\}$\@.
Interestingly, by selecting small $q_2$, for peripheral events we can even achieve a negative $\tilde v_2\{2\}$\@.
When selecting $q_3$ to be small we even see a negative $\tilde v_3\{2\}$ for all centralities.

One can also look at the difference between $v_n\{2\}$ and $v_n\{4\}$, to see whether that changes when using ESE\@. 
For the $v_n\{k\}$ measurements there are two main sources of fluctuations. This is because firstly the QGP itself fluctuates on an event-by-event basis and secondly each event produces only a finite number of thermally sampled particles, which also causes fluctuations.
For high multiplicity the latter contribution is small and it follows from \eqref{eq:Qnv2} (recall $v_n\{2\} \propto \sqrt{\langle q_n^2 \rangle}$) that
\begin{align}
v_n\{2\} & = (\langle v_n\rangle^2 + \sigma_{v_n}^2)^{1/2} \simeq \langle v_n\rangle + \frac{1}{2}\frac{\sigma_{v_n}}{\langle v_n\rangle}, \\
v_n\{4\} & = (\langle v_n\rangle^4 - 2\langle v_n\rangle^2\sigma_{v_n}^2 - \sigma_{v_n}^4)^{1/4} \\
& \simeq \langle v_n\rangle - \frac{1}{2}\frac{\sigma_{v_n}}{\langle v_n\rangle},\nonumber
\end{align}
where $\langle v_n\rangle$ is the mean of the distribution from which the underlying event $v_n$ are taken, and $\sigma_{v_n}$ is its standard deviation \cite{0708.0800,0809.2949}\@.
This means that the difference between $v_n\{2\}$ and $v_n\{4\}$ is a measure of the width of the distribution of the $v_n$ underlying the events.
In Fig.~\ref{fig:observables}, one can see that if one selects for either very large $q_2$ or very small $q_2$, the resulting $v_2\{2\}$ and $v_2\{4\}$ are very close together, indicating that the $v_2$ underlying the events are also very close together, and that we are really taking a small slice of underlying event shapes by selecting on $q_2$\@.
One can similarly see that when selecting either very large or very small $q_3$, the differences between $v_3\{2\}$ and $v_3\{4\}$ become small.

Interestingly, from ESE we can also obtain information that goes beyond just mean $p_T$, $\tilde v_2$ and $\tilde v_3$, as we can also obtain correlations between these quantities.
One can for example observe that for central collisions $\langle p_T\rangle$ is larger for the 90--100\% $q_2$ class than for the 0--10\% $q_2$ class.
This indicates that for central collisions $\langle p_T\rangle$ is positively correlated with $q_2$, and hence with $v_2$\@.
We can also similarly see that this ordering reverses at around 50\% centrality, indicating a negative correlation for more peripheral collisions.
This is in good qualitative agreement with \emph{Trajectum} predictions of the $\rho(v_2\{2\}^2,\langle p_T\rangle)$ observable, which measures this correlation directly \cite{1601.04513, 2004.14463}\@.
Note however that $\rho(v_2\{2\}^2,\langle p_T\rangle)$ does not change sign in experiments \cite{1907.05176}\@.

Another correlation that is visible in Fig.~\ref{fig:observables} is that between $v_2$ and $v_3$\@.
One can see that $v_3\{2\}$ is smaller when selecting the largest values of $q_2$, indicating that $v_3\{2\}$ is negatively correlated with $q_2$, and hence with $v_2$\@.
One can see a similar effect on $v_2\{2\}$ when selecting on $q_3$\@.
As in the previous case, this observation can be corroborated using a previously measured observable, in this case the symmetric cumulant $SC(3, 2)$, which measures the correlation between $v_2$ and $v_3$, and is negative both for our MAP parameters and in experiment \cite{1604.07663}\@.

\begin{figure}[t!]
\includegraphics[width=0.39\textwidth]{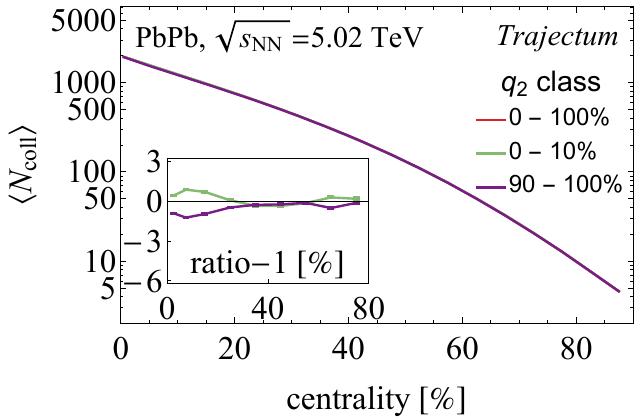}
\caption{\label{fig:ncoll}We show average number of binary collisions $\langle N_{\rm coll}\rangle$ as a function of centrality and using $q_2$ ESE selection. For central and very peripheral events it can be seen that anisotropic geometries lead to a reduction of $\langle N_{\rm coll}\rangle$ by about 2\% and 3\% respectively as compared events without ESE.}
\end{figure}

Finally, in Fig.~\ref{fig:ncoll} we show average number of binary collisions $\langle N_{\rm coll}\rangle$ for ESE events. An increased number of binary collisions is directly related to an increased number of hard scatterings. In \cite{1507.06194} it was shown that this can vary considerably depending on ESE cuts. For our set-up the effects are modest (around 2\%), but we note that for this it is essential for a centrality class to average over smaller subclasses no larger than 1\% in size.

\section{Conclusions and Outlook}
In this work, we studied several different measures of the path length that a hard probe experiences while traversing the QGP\@. These measures should have an important contribution to the amount of energy lost by such a probe. We subsequently studied how differences in these path lengths can be brought out by a judicious choice of cuts on the events and the probes themselves.
In particular, we examined the effect of event shape engineering, as well as a selection of hard probes which are in-plane as opposed to out-of-plane.

By itself, ESE does not modify the average path lengths of non-interacting probes traversing the medium. However, selecting probes which are in-plane yields a significantly shorter path length than selecting probes which are out-of-plane, especially when taking into account that in-plane probes often experience significant fluid flow in the direction of their path.
Furthermore, these differences can be made larger (by about a factor 2) by making tighter $\Delta\phi$ cuts on which probes are considered in-plane and out-of-plane.
By selecting the most elliptical events ESE further increases this difference by another factor of 2, which can yield effective path length ratios $\langle L_{\text{out-of-plane}} \rangle / \langle L_{\text{in-plane}} \rangle$ of up to 2.5 for peripheral collisions or about 1.1 for ultracentral events. Curiously, we did not observe any path length effects within our statistical precision (less than 0.1\%) when using $q_3$ classes as opposed to $q_2$ classes.

It is these large factors that can make ESE valuable to study jet energy loss experimentally. Specifically, it should be possible to study nuclear modification factors according to the $q_2$ and emission angular classes as shown in Fig.~\ref{fig:inoutratio}, whereby in particular depending on centrality the effect of energy loss on the nuclear modification factor should be almost 2.5 times larger in the out-of-plane direction \footnote{A subtle but important point is that even without energy loss partons are significantly modified by nuclear parton distribution functions (see e.g.~\cite{2007.13758})\@. For central to moderately peripheral PbPb collisions the medium energy loss should however be dominant and we can ignore this issue.}\@. Here we assume that the flow-temperature-weighted path length is a good proxy for the average energy loss. This could give a rather clean interpretation of the similar measurement of measuring $v_2$ of high $p_T$ hadrons or jets \cite{2111.06606,1509.07334}, though we note that in those references no ESE was attempted. Incidentally, our Fig.~\ref{fig:inoutratio} shows that fluid flow in energy loss studies contributes a large effect, that can be important for the consistent underprediction of the high $v_2$ at large $p_T$ found in many models \cite{1602.03788, 1902.03231}\@.

We also studied back-to-back path lengths and examine the ratio of the shortest over the longest path length.
The average of ratios thus obtained is between 0.4 and 0.5 for events of up to about 50\% centrality, meaning that dijets can be used to study probes of different path lengths.
Interestingly, the average ratio can be pushed lower by selecting only back-to-back paths which are produced in-plane, and only selecting events in the 90--100\% $q_2$ percentile.

We hence conclude that not only dijets can be used experimentally to study parton energy loss, but also single jets as long as a distinction is made between in-plane and out-of-plane jets.
Furthermore, the distinction between in-plane and out-of-plane dijets is also a useful tool to enlarge path length differences, and for both single jets and dijets ESE can subsequently be used to further increase these differences. 

ESE is not just useful to study hard probes, but also opens up a large number of possibilities for observables in the soft sector.
It is for example possible to study correlations between observables such as $\langle p_T\rangle$ and $v_n\{k\}$\@.
These correlations are consistent with observables such as $\rho(v_n\{2\}^2,\langle p_T\rangle)$ and $SC(m, n)$, but may offer statistically easier methods to include such correlations, potentially allowing their inclusion in Bayesian analyses.
Such estimates would not include the careful bias subtractions that are included in the $\rho(v_n\{2\}^2,\langle p_T\rangle)$ observable, but would nevertheless contain similar information about the initial state which could be useful in a comparison between theory and experiment.

In the future, one of the most immediate avenues of research pointed to by this work is to use in-plane vs.~out-of-plane selection in combination with ESE in experimental studies of hard probes, so that differences in path length can be studied to their fullest extent. 
In addition, in terms of theoretical developments it would be useful to follow this work up with a simulation of a full parton shower in \emph{Trajectum}, which should offer a more realistic model for energy loss than the relatively simple proxies used in this work. %

\section*{Acknowledgements}
We thank Jasmine Brewer, Hannah Bossi, Helen Caines, Cvetan Cheshkov, Aleksas Mazeliauskas and Guilherme Milhano for discussions, and in particular Krishna Rajagopal for a careful reading of the manuscript.
This work was supported in part by the Office of Nuclear Physics of the U.S. Department of Energy. CB and MS are supported by the US DOE under award number DE-SC004168\@.
GN is supported by the U.S. Department of Energy, Office of Science, Office of Nuclear Physics under grant Contract Number DE-SC0011090\@.

\bibliography{Paper,Manual}
\end{document}